# Magnetic Circular Dichroism Measurements of Thin Films

Wala Dizayee[1], Minju Ying[1,2], A. Mark Fox[1] and Gillian A. Gehring[1]

[1]Department of Physics and Astronomy, Hicks Building, University of Sheffield, Sheffield S3 7RH, United Kingdom

[2]Key Laboratory of Beam Technology and Material Modification of Ministry of Education, College of Nuclear Science and Technology, Beijing Normal University, Beijing 100875, P.R. China

**Abstract**

The difference in the transmission for left and right circularly polarised light though thin films on substrates in a magnetic field is used to obtain the magnetic circular dichroism of the film. However there are reflections at all the interfaces and these are also different for the two polarisations and generate the polar Kerr signal. In this paper the contribution to the differences to the total transmission from the transmission across interfaces as well as the differences in absorption in the film and the substrate are calculated. This gives a guide to when it is necessary to evaluate these corrections in order to obtain the real MCD from a measure of the differential transmission due to differential absorption in the film.

## 1. Introduction

Magnetic circular dichroism is given by the difference in transmission for left and right circularly polarised light though a film that is magnetised in the direction normal to the surface. It is a very powerful technique because the MCD spectrum gives information about the magnetic polarisation of the states of the solid as a function of their energy [1-3]. In doped ZnO samples we have seen that there is a strong contribution from the states near the band edge [1]. In cobalt doped $In_2O_3$ there is a positive contribution from transitions to the states due to singly charged oxygen vacancies in the gap [2]. It has also been possible to track the change in the band states on $Zn_{1-x}Co_xO$ over the whole range of concentration, $x$ [3] and to determine the fraction of the whole sample that is in the form of metallic cobalt in ZnCoO [4]. It is also a very sensitive technique to determine weak absorption in a non-magnetic material in a magnetic field when a direct measure of the absorption cannot be separated from surface scattering [5,6].

In a material that is not birefringent and is in the presence of an applied magnetic field the normal modes of light propagation are circularly polarised corresponding to refractive indices are $n_\pm$. For a film of thickness $d$ the MCD|$_{film}$ varies as Im($n_+ - n_-$)$d$ and the Faraday rotation, $\theta$|$_{film}$ as Re($n_+ - n_-$)$d$ as these effects depend on the transmission of the light through the body of the film. In a transparent material of refractive index $n$ there is little absorption of any kind hence Im($n_+ - n_-$) is usually considerably smaller than Re($n_+ - n_-$). The effect of the change in transmission due to the polar Kerr effect which is also present in these samples at each interface should be considered in these studies. In this paper we demonstrate the effects of including the differential transmission due the changes in the real part of the refractive index on the accurate determination of the MCD.

In Section 2 we evaluate the total differential transmission in terms of the transmission within the film and the substrate and the transmission across all the interfaces in terms of the real and imaginary parts the refractive indices for circular polarisation within the film and the substrate. The effects of multiple passes through the system are also discussed briefly. The importance of including interface effects is shown to be dependent on the film thickness, $d$. The theory is illustrated for some magnetic and nonmagnetic films in Section 3. Finally the conclusions are in Section 4.

## 2. Calculation of the total differential transmission of a thin film on a substrate

The calculation will be done assuming a single pass. The theory can be generalised to allow for interference in the film and multiple passes through the substrate [7] but this adds greatly to the complexity and is not physically correct if the substrate is not flat. The transmission from a medium with refractive index $n_1$ into a medium with index $n_2$ is given by $T_{1,2} = \dfrac{n_1 n_2}{(n_1 + n_2)^2}$. We shall need to consider three interfaces: the interface between the air, or vacuum, and the film, the interface between the film and the substrate and the interface between the substrate and the air. In the very weakly absorbing materials that concern us here we can use the real values of $n_1$ and $n_2$ here. This means that there will be contributions to the differential transmission across the interfaces that depend on the Faraday rotations for the film and substrate but not on the film thickness. The measured value of the difference in the total transmissions, $\Delta T = \dfrac{T_+ - T_-}{T_+ + T_-}$, depends both on the effect of the difference in transmission within the film and the substrate but also on the transmission across the interfaces. In this paper we calculate the contribution from the interface transmission to investigate the magnitude of this effect and to determine the range of film thicknesses for which this effect is likely to be important.

We consider the situation where the film is sufficiently rough that multiple reflections from the back surface of the substrate and the front surface of the film may be neglected. The corrections may be shown to be insignificant [7,8]. The film of thickness $d$ and refractive index $n$ is on a substrate of thickness $L$ with refractive index $s$.

The total transmission is given in terms of the transmission factors at the front, the film-air interface, $T_f = \frac{4n}{(n+1)^2}$, the film–substrate interface $T_{fs} = \frac{4ns}{(n+s)^2}$ and at the back, the substrate-air interface, $T_b = \frac{4s}{(s+1)^2}$, and the absorption that occurs in the film, $e^{-\alpha d}$, and in the substrate $e^{-\gamma L}$.

$$T_{total} = \frac{64 n^2 s^2 e^{-\alpha d} e^{-\gamma L}}{(n+1)^2 (n+s)^2 (s+1)^2} \qquad [1]$$

In a magnetic field all the coefficients, $\alpha, \gamma, n, s$ take different values for left and right circularly polarised light. The film and substrate are weakly absorbing and hence we can ignore the imaginary parts of $n$ and $s$ in this equation but, of course, include the imaginary part of $n$ in $\alpha$ and the imaginary part of $s$ in $\gamma$.

The MCD is related to the difference in transmission due to the absorption for left and right circularly polarised light [9].

$$MCD = \frac{1}{2}\left[\frac{T^+ - T^-}{T^0}\right] \qquad T_0 = \frac{T^+ + T^-}{T_0} \qquad [2]$$

Magnetic effects are very weak so we can work to first order in the magnetic effects. This leads to the following expression for $\frac{T^+_{total} - T^-_{total}}{T^0_{total}}$;

$$\frac{T^+_{total} - T^-_{total}}{T^0_{total}} = \left[ 2MCD|_{film} + 2MCD|_{substrate} \right]$$

$$+ \frac{(n_0-1)\operatorname{Re}(n^+ - n^-)}{n_0(n_0+1)} + \frac{(s_0-1)\operatorname{Re}(s^+ - s^-)}{s_0(s_0+1)} + \frac{(n_0-s_0)\operatorname{Re}(n^+ - n^-)}{n_0(n_0+s_0)} + \frac{(s_0-n_0)\operatorname{Re}(s^+ - s^-)}{s_0(n_0+s_0)}$$

[3]

This contains the term that we seek to measure MCD|$_{film}$ and a term related to the MCD|$_{substrate}$ of the substrate and also terms that depend on the field dependences of the real parts of the refractive indices $n$ and $s$ which may be found using Faraday measurements.

The contributions of the film alone may be found by subtracting the differential transmission from a bare substrate which is given by,

$$\frac{T^+_{sub} - T^-_{sub}}{T^0_{sub}} = 2MCD|_{substrate} + \frac{2(s_0-1)\operatorname{Re}(s^+ - s^-)}{s_0(s_0+1)}.$$

Finally we obtain the following expression for the magneto-optical contributions that depend on the film,

$$\frac{T^+_{total} - T^-_{total}}{T^0_{total}} - \frac{T^+_{sub} - T^-_{sub}}{T^0_{sub}} = 2MCD|_{film}$$

$$+ \frac{(n_0-1)\operatorname{Re}(n^+ - n^-)}{n_0(n_0+1)} - \frac{(s_0-1)\operatorname{Re}(s^+ - s^-)}{s_0(s_0+1)} + \frac{(n_0-s_0)\operatorname{Re}(n^+ - n^-)}{n_0(n_0+s_0)} + \frac{(s_0-n_0)\operatorname{Re}(s^+ - s^-)}{s_0(n_0+s_0)}$$

[4]

We see that the first term in [4] depends on the thicknesses of the film; this gives the real MCD|$_{film}$. The other terms are independent of thickness and depend on the difference in reflection for left and right circularly polarised light at the interfaces. Thus these terms will be negligible for thick films.

The value of MCD|$_{film}$ is given in terms of the difference between the imaginary parts of the refractive indices $n^\pm$,

$$MCD|_{film} = \frac{\pi d}{\lambda} \operatorname{Im}(n^+ - n^-) .$$ [5]

## 3. Evaluation of the MCD of a thin film from the total differential transmission

In this section we consider how to relate the various terms in equation [4] to magneto-optic quantities, the MCD and Faraday effects of the film and the substrate.

The terms that involve the real parts of the refractive indices may be related to the measured Faraday effects of the film and the substrate [9],

$$\theta_{film} = -\frac{\pi d}{\lambda}\text{Re}(n^+ - n^-), \qquad \theta_{substrate} = -\frac{\pi L}{\lambda}\text{Re}(s^+ - s^-) \qquad [6]$$

$$\frac{T^+_{total} - T^-_{total}}{T^0_{total}} - \frac{T^+_{sub} - T^-_{sub}}{T^0_{sub}} = 2MCD|_{film} - \frac{(n_0-1)\lambda\theta_{film}}{n_0(n_0+1)\pi d} + \frac{(s_0-1)\lambda\theta_{substrate}}{s_0(s_0+1)\pi L} - \frac{(n_0-s_0)\lambda\theta_{film}}{n_0(n_0+s_0)\pi d} - \frac{(s_0-n_0)\lambda\theta_{substrate}}{s_0(n_0+s_0)\pi L} \qquad [7]$$

The importance of this correction may be assessed roughly by studying the ratio of the first two terms in equation (4) because the contribution from the substrate is usually extremely small. Using equation (5) this ratio can be written as,

$$\frac{\text{Im}(n^+ - n^-)n_0(n_0+1)2\pi d}{\text{Re}(n^+ - n^-)(n_0-1)\lambda} = \frac{MCD|_{film}}{\theta_{film}} \frac{n_0(n_0+1)2\pi d}{(n_0-1)\lambda} \qquad [8]$$

The factor $\frac{2\pi n_0(n_0+1)}{(n_0-1)}$ will be considerable larger than unity and so the correction will be important only if the MCD is particularly small compared with the Faraday rotation or the thickness is such that $d<<\lambda$.

We consider the magnitude of the correction for two examples: a nonmagnetic oxide film of $MoO_x$, and a magnetic film of ZnCo both on sapphire substrates.

The Faraday rotation (FR) should be measured for the substrate alone and then the film plus the substrate, the Faraday effect for the film is obtained by subtracting these two measurements. We note that in equation [6] the FR from the film and the substrate are divided by the thickness of the film and the substrate respectively. The expressions given in equations (2), (5) and (6) are in radians, we convert these to degrees in accord with common practice.

We measure the effects of including the corrections for two examples: a nonmagnetic oxide film of $MoO_x$, of thickness 50nm and a magnetic film of $Zn_{0.9}Co_{.1}O$ of thickness 137nm on sapphire substrates.

In figure 1 we show the measured Faraday rotation per unit thickness for the thin films and the substrate, the signal from the substrate obviously much smaller than for the films and may be neglected in most cases. The plots of $Re(n^+ - n^-)$ are also given and we note that the strong energy dependence observed in the Faraday rotations has been partially cancelled out by the inclusion of the wavelength factor, $\lambda$. This occurs physically because $Re(n_+ - n_-)$ has contributions from electronic transitions to all the excited energy levels and so is relatively independent of energy unlike the MCD which depends directly polarisation of the electronic states involved in the allowed energy transitions. This means that the correction, even when sizeable, changes the magnitude of the MCD but without changing its shape in any significant way. The magnitude of $Re(s^+ - s^-)$ is much smaller than those for the films and so this term does not contribute significantly to the correction factor.

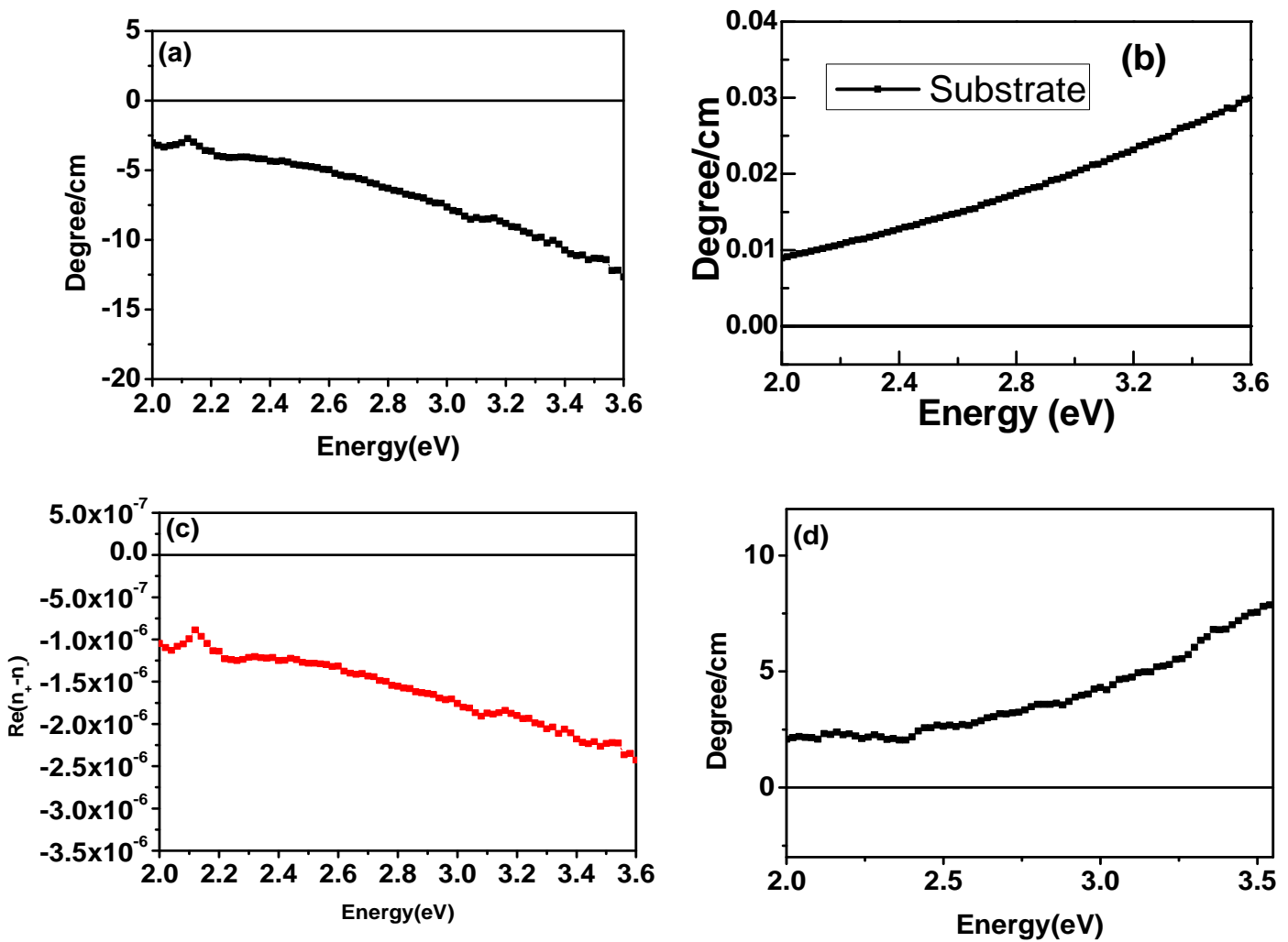

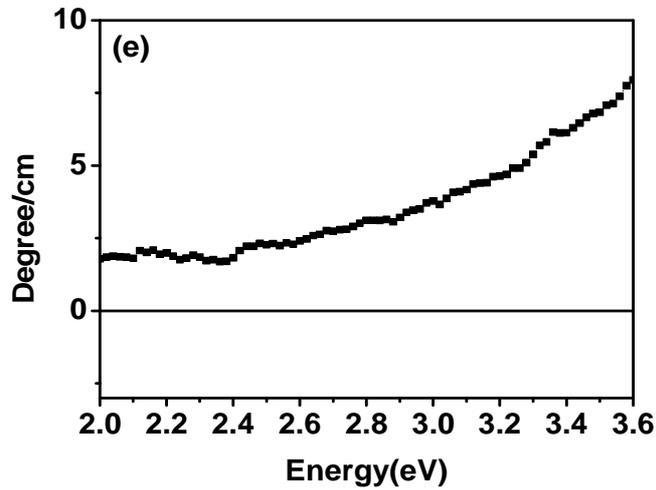

Fig1. (a) The Faraday rotation of films of $MoO_x$ where the contribution from the sapphire substrate has been subtracted; (b): the Faraday rotation for the bare sapphire substrate (c) The plots of $Re(n_+ - n_-)$ (d) the apparent MCD of films of $MoO_x$ after subtraction of the MCD of the bare substrate but the effects of the Kerr reflections have not been made (e) The corrected values for the MCD.

The results for $MoO_x$ are presented in figure 1. These include the Faraday rotation per unit thickness (FR) of the film, after subtracting the contribution of the substrate and the FR of the substrate. The FR signal from the substrate obviously is much smaller than for the film and may be neglected in most cases. The value of $Re(n^+ - n^-)$ is obtained from $\lambda\theta_{film}$. We see that the strong energy dependence observed in the Faraday rotations has been almost cancelled out by the inclusion of the factor of $\lambda$. This occurs physically because $Re(n_+ - n_-)$ has contributions from electronic transitions to all the excited energy levels and so is relatively independent of energy. This means that the correction, even when sizeable, change the magnitude of the MCD but without changing its shape in any dramatic way. The apparent MCD of the film and the corrected value are shown in figs (d) and (e). It is seen that the correction is small but finite even with a film as thin as 50nm. This is due to the fact that the magnitude of the FR is not that much larger than that of the MCD.

Figure 2 includes all the comparable results for the ZnCoO film. The change introduced by including the interface effects is even smaller here as the the film is thicker, 137nm compared with 50nm for the $MoO_x$.

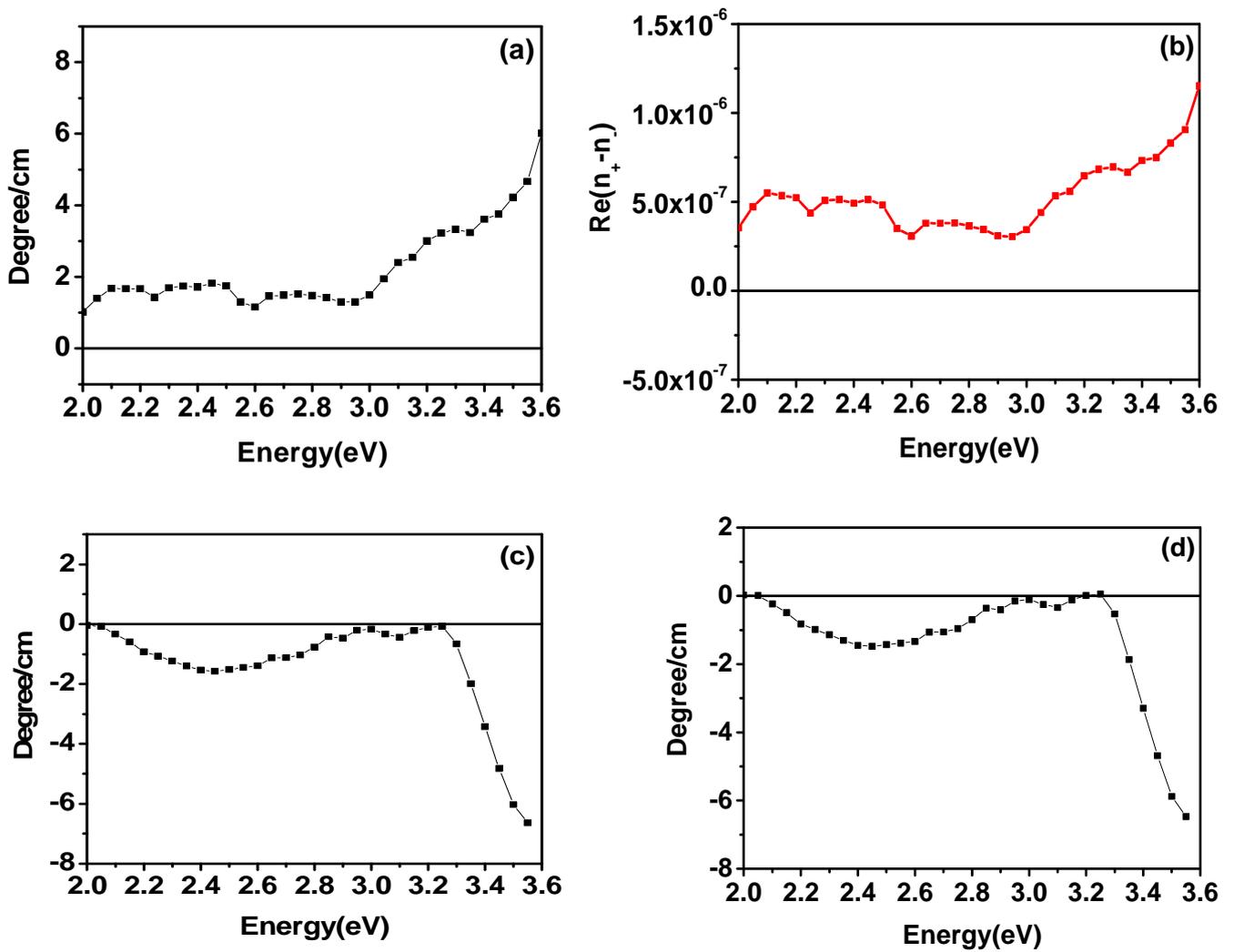

Fig2. (a) The Faraday rotation of ZnCoO film where the contribution from the sapphire substrate has been subtracted (b) The plots of $Re(n_+ - n_-)$ (c) the apparent MCD of films of ZnCoO after subtraction of the MCD of the bare substrate but the effects of the Kerr reflections have not been made (d) The corrected values for the MCD

## 4. Conclusion

The corrections that should be applied to a measure of the observed difference in the transmission for left and right circularly polarised light to obtain the MCD due to differential absorption in the film has been evaluated. These corrections arise from the differential transmission across all the interfaces in the film/substrate structure. It was found that the corrections can be evaluated by using measurements of the Faraday rotations of the film and the substrate. The condition for the corrections to be significant were that the films were very thin and moreover the MCD was relatively small compared with the Faraday rotation of the film. The corrections depend on the real part of the refractive indices and so are relatively smooth functions of energy. Hence these corrections, even when large, do not change the shape of the MCD spectra significantly.